\documentclass[structabstract]{aa}  
\usepackage{graphicx}
\usepackage{sidecap}
\usepackage{natbib}

\usepackage{savesym}
\usepackage{txfonts}
\savesymbol{iint}
\savesymbol{iiint}
\savesymbol{iiiint}
\savesymbol{idotsint}

\usepackage[fleqn]{amsmath}
\restoresymbol{ams}{iint}
\restoresymbol{ams}{iiint}
\restoresymbol{ams}{iiiint}
\restoresymbol{ams}{idotsint}

\newcommand{\lsim}{\mathrel{\hbox{\rlap{\lower.55ex \hbox{$\sim$}} \kern-.3em \raise.4ex \hbox{$<$}}}}

\usepackage{tabularx}
\newcolumntype{L}[1]{>{\raggedright\arraybackslash}p{#1}} 
\newcolumntype{C}[1]{>{\centering\arraybackslash}p{#1}} 
\newcolumntype{R}[1]{>{\raggedleft\arraybackslash}p{#1}} 

\bibpunct{(}{)}{;}{a}{}{,} 

\begin{document}
   \title{HST observations of the limb polarization of Titan}


   \author{A. Bazzon\inst{1}, H.M. Schmid\inst{1}, E. Buenzli\inst{2}}

   \institute{ETH Zurich, Institute of Astronomy,
              Wolfgang-Pauli-Str. 27, 8093 Zurich, Switzerland
              \and
   	     Max-Planck-Institut f{\"u}r Astronomie,
   		K{\"o}nigstuhl 17, 69117 Heidelberg, Germany\\
              \email{bazzon@astro.phys.ethz.ch} }

   \date{Draft version}

\authorrunning{A. Bazzon, H.M. Schmid, E. Buenzli}

 
  \abstract
   {Titan is an excellent test case for detailed studies of the scattering polarization from thick hazy atmospheres. 
   Accurate scattering and polarization parameters have been provided by the in situ measurements
   of the Cassini-Huygens landing probe. For Earth-bound observations Titan can only be observed at a backscattering situation, where the disk-integrated
   polarization is close to zero. However, with resolved imaging polarimetry a second order polarization signal along the entire limb of Titan can be measured.
 }
   {We present the first limb polarization measurements of Titan, which are compared as a test to our limb polarization models.
   }
   {Previously unpublished imaging polarimetry from the HST archive is presented which resolves the disk of Titan. 
   We determine flux-weighted averages of the limb polarization and
    radial limb polarization profiles, and investigate the degradation and 
   cancelation effects in the polarization signal due to the limited 
   spatial resolution of our observations. Taking this into
   account we derive corrected values for the limb polarization in Titan. The results are compared with limb polarization models, using atmosphere and haze
   scattering parameters from the literature.
    }
   {In the wavelength bands between 250 nm and 2 $\mu$m a strong limb polarization 
    of about $2-7~\%$  is detected with a position angle perpendicular to the
    limb. The fractional polarization is highest around 1~$\mu$m.
    As a first approximation, the polarization seems to be equally strong along 
    the entire limb.
   The comparison of our data with model calculations and the literature shows that the detected polarization is compatible with 
    expectations from previous polarimetric observations taken with Voyager 2, Pioneer 11, and
    the Huygens probe.
  }
   {Our results indicate that ground-based monitoring measurements of the limb-polarization of Titan 
   could be useful for investigating local haze properties and the impact of short-term and seasonal variations of the hazy atmosphere of Titan.
   Planets with hazy atmospheres similar to Titan are particularly good candidates for detection with the polarimetric mode of the
   upcoming planet finder instrument at the VLT.  Therefore, a good knowledge of the polarization properties of Titan is also important for the search and 
   investigation of extra-solar planets.
}


   \maketitle
%

\section{Introduction}

Solar light reflected from planets, moons, and smaller objects is polarized. 
This basic property of light reflection provides a powerful diagnostic 
tool for the remote investigation of the scattering particles in 
atmospheres and the reflecting surfaces of solar system bodies. 

Polarization studies of Titan are particularly well suited 
to studying the scattering polarization from a hazy atmosphere. 
The hazy atmosphere of Titan produces a very strong polarization signal over a wide
wavelength range. 
At quadrature phase
$\alpha \approx 90^\circ$, the fractional polarization of Titan from the UV to the $R$ band is $p\approx 50~\%$, as measured by the
Pioneer 11 \citep{Tomasko1982} and \mbox{Voyager 2} \citep{West1983} spacecrafts.
Furthermore, thanks to the joint NASA-ESA Cassini-Huygens satellite mission, Titan's surface
and atmospheric structure are known in great detail. In particular, the measurements made inside
Titan's atmosphere, made available by the Huygens landing probe, have provided accurate scattering and polarization parameters for the haze particles
(e.g., \citealt{Brown2010} and \citealt{Tomasko2008}).

For Earth-bound instruments the atmosphere of Titan can only be observed at very limited phase angles $\alpha\lesssim 5^\circ$, where the disk-integrated
polarization is close to zero, and the polarimetric properties can only be investigated if disk-resolved imaging polarimetry is available.
In this work we present previously unpublished
imaging polarimetry from the HST archive which
resolves the disk of Titan and clearly shows a strong limb
polarization effect. 
Together with literature values for the albedo $A_{\rm g}$ and the quadrature polarization $p(90^\circ)$ of Titan, our limb polarization measurements can now be used to
test polarization models for a haze scattering atmosphere, and we can make predictions for the detection and characterization of reflected light of extra-solar planets. 

The limb polarization
is a well-known second order scattering effect of reflecting
atmospheres with predominantly Rayleigh-type scattering processes
\citep[e.g.,][]{VandeHulst1980}.
In general, single backscattering with scattering angles $\sim$$180^\circ$ would produce a very small polarization signal or no signal at all. 
Thus, the polarization measured at the limb arises from second order and also higher order scatterings by light that is 
scattered sideways, i.e., more or less parallel to the limb, 
and then scattered back to the observer. 
The polarization angle induced by Rayleigh scattering, i.e single dipole-type scattering, is perpendicular to the 
propagation direction of the incoming photon.
Hence, the position angle of polarization is perpendicular to the limb everywhere. 

Over the last 20 years, Titan's thick and hazy atmosphere has been monitored and intensively studied by spectral HST observations,
revealing strong local albedo variations mainly caused by seasonal migration of haze from one hemisphere to another \citep[][and references therein]{Lorenz2004}. 
The most prominent features are a varying north-south asymmetry, a dark polar hood that is most prominent in the UV, and a detached haze layer
lying only $\sim$200 km above the optical limb of Titan \citep[e.g.,][]{Lorenz2004, Lorenz2006}.
Observations of Titan from the ground, e.g., using the upcoming SPHERE instrument at the VLT \citep{Beuzit2008}, 
consisting of state-of-the-art imagers and polarimeters, will have the
advantage of a much higher spatial resolution and a broader wavelength coverage. Therefore, a consistent monitoring program of Titan's atmosphere
from the ground could potentially be useful for investigating the temporal changes in the polarization structure along the limb, and for constraining local haze
properties of Titan.

After the description of the observational data and the basic data reduction in Sects. \ref{observations} and \ref{data reduction} respectively, we
discuss the intensity images in Sect. \ref{s:intensity} and compare them with the literature.  
In Sect. \ref{sectqu} we derive the Stokes images for all our wavelength bands, which are 
then converted into radial limb polarization images in Sect. \ref{s:radial polarization}.
There, we discuss the radial polarization profiles and the advantage of disk-integrated radial polarization measurements, and we explain how
we correct our images for the polarization degradation caused by a PSF smearing effect.
In Sect. \ref{modelsect} we describe our radiative transfer model for Titan, and we compare the model with our limb polarization results and 
literature values for the geometric albedo $A_{\rm g}$ and the quadrature polarization $p(90^\circ)$ of Titan.
The last section gives a summary and discusses the prospects for a polarimetric monitoring program of Titan and for the detection and investigation of extra-solar haze planets.


\section{Observations}\label{observations}

We reduced and analyzed imaging polarimetry of Titan from the HST
archive\footnote{HST proposal ID 9385} for which only the intensity images have been published \citep{Lorenz2006} but not the polarization data. 
The data were recorded with the ACS HRC
and the NICMOS instruments in seven filters covering wavelengths 0.25 $\mu$m - 2 $\mu$m. 
Polarimetry is achieved with three subsequent measurements, using three polarizers with different orientations.
Titan was observed in 2002 during two visits on November 27 and December 2, i.e., shortly after
southern summer solstice that occurred in late October 2002, and with the north pole of Titan on the hidden hemisphere.
Table \ref{log} gives an overview of the observational parameters, the used instruments and corresponding filters, the total exposure times,
and the plate scales. 

For broad-band observations
the nominal filter wavelength may differ significantly from 
the average wavelength of the photons registered in 
the polarization map. Similarly the
evaluation of the reflected-flux weighted albedo $A_{\rm g,eff}$ for a given filter considers
a weighting with the effective spectral distribution of the registered photons, taking into account the
wavelength dependence of the instrument, the solar
photon spectrum, and the albedo of Titan. 

     \begin{figure}
        \centering
        \includegraphics[width=8.8cm]{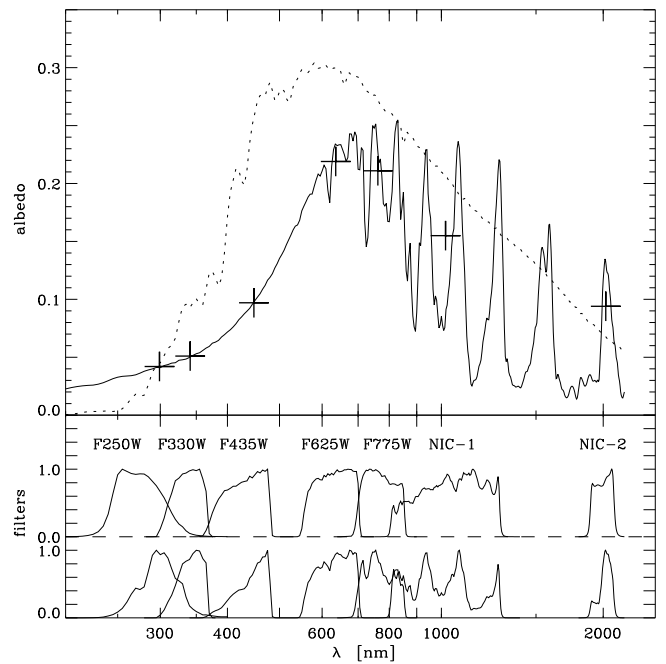}
                 \caption{The geometric albedo of Titan (upper panel) is given by the
		full line and the crosses indicate the effective albedos $A_{\rm g,eff}$ and wavelengths $\lambda_{\rm eff}$ for
		the HST filter polarimetry. The dashed line illustrates the solar spectrum. 
		The lower panel shows the normalized filter efficiency curves and the 
		normalized wavelength distribution of the registered
		photons.
                 }
                 \label{figalbedo}         
        \end{figure}

Figure \ref{figalbedo} illustrates the spectral dependence of the full disk
albedo of Titan from the mid-UV to the near-IR
(\citealt{McGrath1998} for the UV,
\citealt{Karkoschka1998} for the visual, \citealt{Negrao2006}
for the near-IR) assuming Titan's optical radius
varying with $\lambda$ according to the \citet{Toon1992} radius $R_{\rm T}$. 
The solar irradiance photon spectrum derived from \citet{Thuillier2004} is given by the dashed line.
The normalized instrument 
efficiencies in the different HST filters and the calculated
spectral distribution of the registered photons are also indicated. 
We use
the median wavelengths of the individual 
photon distributions as effective filter wavelengths $\lambda_{\rm eff}$,
and the reflected-flux weighted values for the 
effective albedos $A_{\rm g,eff}$ for each filter. The corresponding
values are indicated in Fig.~\ref{figalbedo} and Table~\ref{intpol} 
lists the derived values and $R_{\rm T}(\lambda_{\rm eff})$. 
   
We note that when looking specifically at the limb the methane bands are weaker and center-to-limb differences in the albedo spectrum are present 
\citep[e.g.,][]{Smith1996, Lorenz2004, Lorenz2006}.  However, we mainly focus on the derivation of disk-integrated albedo and limb polarization values (Sect.~\ref{s: disk-integrated radial polarization}).

The effective filter wavelengths $\lambda_{\rm eff}$ are shifted to longer wavelength
for the UV/blue filters because of the steep photon spectrum. The effective albedos $A_{\rm g,eff}$ are relevant for the
filters covering spectral regions with strong CH$_4$ absorption bands.
They are about 20~\% higher for the $NIC1$ and
40~\% higher for the $NIC2$ passbands than the simple mean. For
the $F250W$ and the $F330W$ filters the difference is $<5~\%$, and for the other filters the difference is $\sim$$10~\%$ or less
(steep flux gradient combined with systematic albedo gradient).

\begin{table}[t!]
\caption{Observational parameters and summary of the HST polarimetry used in this work. The southern summer solstice of Titan occurred in late October 2002. 
NP distance gives the angular distance of the north pole from the center of the disk, whereas the negative distance indicates that the north pole is on the hidden hemisphere.}
\label{log}
\centering
\begin{tabular}{lcc}
\hline \hline
                   &  ACS HRC & NICMOS 1\&2  \\
\hline
obs. parameters$^{\,\rm a,b}$       &            &    \\
\,\,\, date              &      2002-12-02  &  2002-11-27   \\
\,\,\, diameter      &     0.88"     &   0.88"     \\
\,\,\, phase angle  &  $1.8^{\circ}$    &      $2.4^{\circ}$      \\
\,\,\, NP distance        &    -0.39"      &     -0.39"       \\
\,\,\, angle NP-${\rm N}_{\rm cel}$   &     $-5.3^\circ$     &    $-5.3^\circ$    \\
exposures      &       &     \\
\,\,\,  $F250W$     & $3\times 4 \times 365$~sec  &  -  \\
\,\,\,  $F330W$     & $3\times 4 \times 200$~sec  &  -  \\
\,\,\,  $F435W$     & $3\times 4 \times 45$~sec  &    -   \\
\,\,\,  $F625W$     & $3\times 2 \times 12$~sec  &    -  \\
\,\,\,  $F775W$     & $3\times 2 \times 9$~sec  &       - \\
\,\,\,  POL S      &      -              &  $3\times 2 \times 20$~sec     \\
\,\,\,  POL L     &       -              &  $3\times 2 \times 12$~sec      \\
plate scale    &  $0.025''$/pix   & $0.043''$/pix \& $0.075''$/pix  \\
\hline
\multicolumn{3}{l}{\vspace{-0.2cm}}\\
\multicolumn{3}{l}{$^{\rm a}$\citet{Almanach2002}}\\
\multicolumn{3}{l}{$^{\rm b}$http://ssd.jpl.nasa.gov/?horizons}
\end{tabular}
\end{table}


\section{Basic data reduction}\label{data reduction}

The data provided by the HST 
data reduction pipeline are already corrected for bias, dark, flatfield, and image distortion. However, in the case of the ACS data the pipeline
drizzled, combined images showed a strange stripe pattern which was caused by incorrectly set bits in the data quality mask that are used for identifying pixels flagged as cosmic rays by MultiDrizzle\footnote{The MultiDrizzle Handbook, Chap.~5.4.6.3 "Final Products"}. Using the instructions provided by
the stsdas\footnote{Space Telescope Science Data Analysis System} helpdesk we reset the 
dq\_bits and re-run multidrizzle using the standard settings. 

Both the ACS and the NICMOS instrument contain sets of three linear polarizer filters with their relative polarization directions oriented
according to $0^\circ$, $60^\circ$, $120^\circ$ and $0^\circ$, $120^\circ$, $240^\circ$ respectively. In the first data reduction step
the basic pipeline processed images were cut out and aligned to subpixel accuracy of $\pm 0.1~\rm pixel$. Then the images corresponding to the
Stokes parameters $I$, $Q$, and $U$ were calculated and corrected for instrumental polarization.

\subsection{Calculation of Stokes parameters for ACS}
In case of ACS the Stokes parameters are calculated according to:
\begin{displaymath}
I = \left(\frac{2}{3}\right)[i_0 \cdot c_{0}(\lambda) + i_{60} \cdot c_{60}(\lambda) + i_{120} \cdot c_{120}(\lambda)]\, ,
\end{displaymath}
\begin{displaymath}
Q = \left(\frac{2}{3}\right)[2 i_0 \cdot c_{0}(\lambda) - i_{60} \cdot c_{60}(\lambda) - i_{120} \cdot c_{120}(\lambda)] \left(   \frac{T_{\parallel} + T_{\perp} }{T_{\parallel} - T_{\perp}}     \right) \, ,
\end{displaymath}
\begin{displaymath}
U = \left(\frac{2}{\sqrt{3}}\right)[i_{60} \cdot c_{60}(\lambda) - i_{120} \cdot c_{120}(\lambda)] \left(   \frac{T_{\parallel} + T_{\perp} }{T_{\parallel} - T_{\perp}}     \right)\, ,
\end{displaymath}
whereas $i_\ast$ indicate the intensity images corresponding to the three polarizer orientations, $c_\ast$ are corresponding correction 
factors calibrating the polarization zero-point, and the $T_\ast$ parameters correct for polarization cross-talks caused by leakages of the polarizing filters.
The calibration parameters are given in the HST calibration handbook (Figure 5.3, Table 6.3) which we summarize in \mbox{Table \ref{calib ACS}}. 
According to the handbook for the ACS  camera the residual instrumental polarization uncertainty should be at the one-part-in-ten level for
highly polarized sources and at the 1 \% level for weakly polarized targets. 

For the $F250W$ filter no calibration parameters are provided by the HST handbook. However, because of symmetry reasons one can assume that
to first order the fractional polarization at the center of the apparent disk of Titan should be zero:
\begin{displaymath}
(Q/I)_{\rm center} \approx (U/I)_{\rm center} \approx 0 \, .
\end{displaymath}
 
Therefore, by using extrapolated values for $c_\ast$ as starting points, and by minimizing the fractional polarization around the
center of Titan, we determined estimates for the $F250W$ calibration parameters indicated in Table \ref{calib ACS}.
Especially for the integrated radial polarization described in Sect. \ref{s:radial polarization} these calibration parameters produce very 
reasonable results which are consistent with the other wavelengths.

 \begin{table}
\caption{Calibration parameters for ACS polarimetry. }
\label{calib ACS}
\centering
\begin{tabular}{clccccc}
\hline \hline
\multicolumn{2}{c}{Spectral Filter}        &         $c_{0}$  &         $c_{60}$ &         $c_{120}$   &         $T_{\parallel}$    &         $T_{\perp}$\\
\hline
&$F250W$$^{\ast\ast}$     &   1.840     &    1.625      &    1.801      &     0.28       &      0.02       \\
&$F330W$                   &       1.7302      &         1.5302      &        1.6451      &         0.475       &      0.05       \\
&$F435W$                  &       1.6378      &         1.4113      &        1.4762      &         0.525       &      0.02       \\
&$F625W$$^{\ast}$                  &       1.0443      &         0.9788      &        0.9797      &         0.500       &      0.00       \\
&$F775W$                  &       1.0867      &         1.0106      &        1.0442      &         0.650       &      0.00       \\
\hline
        \multicolumn{7}{l}{\vspace{-0.2cm}}\\
        \multicolumn{7}{p{8cm}}{$^\ast$Calibration not scaled for Stokes $I$. 
        According to HST handbook there is also some evidence of a polarization pathology.} \\
        \multicolumn{7}{p{8cm}}{$^{\ast\ast}$No calibration parameters available. See text for the derivation of these values.}
\end{tabular}
\end{table}

\subsection{Calculation of Stokes parameters for NICMOS}\label{NICMOS calibration}
In case of NICMOS the HST handbook provides the user with two coefficient matrices $M_{1;2}$ to calculate the Stokes parameters of $NIC1$ and $NIC2$ respectively:
\begin{displaymath}
   \left(
    \begin{array}{c}
            I\\Q\\U
     \end{array}
      \right)
 = M_{1;2}^{-1}
    \left(
    \begin{array}{c}
            i_{0}\\i_{120}\\i_{240}
     \end{array}
      \right)
\end{displaymath}
with the matrices
\begin{displaymath}
  M_{NIC1} = 
    \left(
    \begin{array}{c c c}
          0.3936   &   0.3820   &  0.0189   \\
          0.3959   &  -0.1118   &  -0.1463   \\
          0.3902   &   -0.2768   &  0.1150   \\
     \end{array}
     \right) \, ,
\end{displaymath}

\begin{displaymath}
  M_{NIC2} = 
    \left(
    \begin{array}{c c c}
          0.5094   &   0.3550   &  0.1131   \\
          0.5139   &  -0.0403   &  -0.3206   \\
          0.5159   &   -0.3262   &  0.3111   \\
     \end{array}
     \right) \, .
\end{displaymath}

For NIC1 the coefficient matrix calibration is not perfect and 
a residual instrumental polarization at a level of $p_{\rm inst.}\approx1.2-1.5~\%$ was reported \citep[see][]{Batcheldor2009}.
Furthermore, for bright targets ghost images are present in two NIC1 polarization filters ($i_{0^\circ},\,i_{240^\circ}$). 
In case of NIC2 the instrumental effects are very well calibrated, and uncertainties as low as $p_{\rm inst.}\approx0.2~\%$ should be achievable with bright objects. 
Both for NIC1 and NIC2, this is in good agreement with our findings for the Titan polarization in the disk center in Section \ref{sectqu}.


\section{Intensity images}\label{s:intensity}

The used HST-dataset was mainly taken to study seasonal
effects of the stratospheric haze on Titan, and an analysis of the spectro-photometric data is given by \citet{Lorenz2004, Lorenz2006}. 
In particular, they describe and explain in detail the varying dark polar hood and the north-south asymmetry measured in different narrow-band filters.  

The dark polar hood was first seen in 1980 by Voyager 1 around the north, and disappeared from the south pole in 2002-2003 \citep{Lorenz2006}. 
Most probably, the polar hood is associated to a downwelling during the long polar night, redistributing
haze from the summer hemisphere towards the winter pole \citep{Rannou2002, Lorenz2006}. This process then also creates the detached haze layer as the haze is 
horizontally drawn from beneath the formation zone.

\citet{Lorenz2004} find that in the narrow-band filters the north-south asymmetry is reversed between the blue (439 nm) and the red (889 nm), and almost absent at 619 nm. 
In the blue Titan is brighter in the south than in the north and vice versa in the red methane bands. In the near-IR, the variation of the asymmetry with wavelength is dramatic, and
different narrow-band filters may see different reversions of the north-south asymmetry as they are probing different altitude regions. 

              \begin{figure}
        \centering
        \includegraphics[width=8cm]{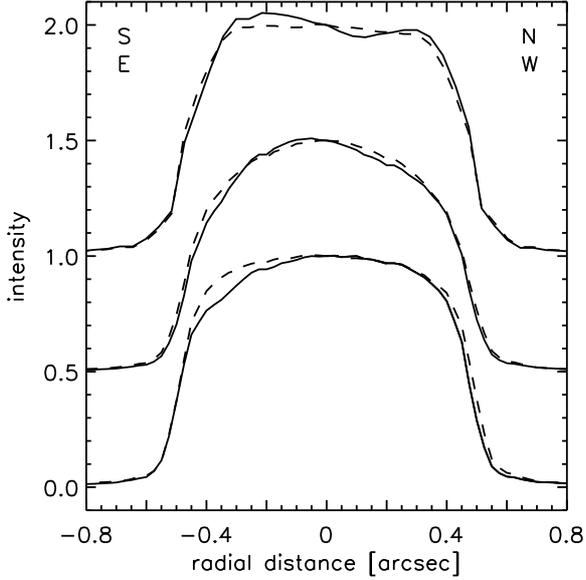}
                 \caption{Intensity cuts/profiles through the disk center for the $F250W$ band (bottom), the $F435W$ band (middle), and the $NIC1$ band (top).
                 The solid line indicates the north-south profile through the planetary poles, and the dashed line is the east-west profile perpendicular to the polar axis respectively.}
                 \label{nsa}    
        \end{figure}

The north-south asymmetry is also visible in our broad-band images ranging from 0.3-2 $\mu$m of the same HST visit as the narrow-band data of \citet{Lorenz2004}, and
it is in qualitative agreement with the spectro-photometric analysis by \citet{Lorenz2004, Lorenz2006}. 
Figure \ref{nsa} shows N-S and E-W profiles of Titan for the $F330W$ band, the $F435W$ band, and the $NIC1$ band. 
Intensity images for the bands $F330W$, $F435W$, $F625W$, $NIC1$, and $NIC2$ are given in the left panel of Fig. \ref{rstokesiq}.
The nominal optical radius of Titan, the south pole, and the equator are also indicated.
The images are normalized such that
\begin{equation}\label{inorm}
\int_0^{R_{\rm int}} \int_0^{2\pi} Ir~drd\phi = A_{\rm g,eff} \cdot \pi R_{\rm int}^2 \,,
\end{equation}
where $A_{\rm g,eff}$ indicates the effective albedo given in Table \ref{intpol} and Fig. \ref{figalbedo}, and $R_{\rm int}= 0.75"$ is the integration radius, which is greater than the
nominal radius of Titan $R_{\rm Titan}=0.44"$.


\section{Stokes $Q$ and $U$ images for Titan}
\label{sectqu}

\begin{figure}
        \centering
        \includegraphics[width=4.1cm,angle=90]{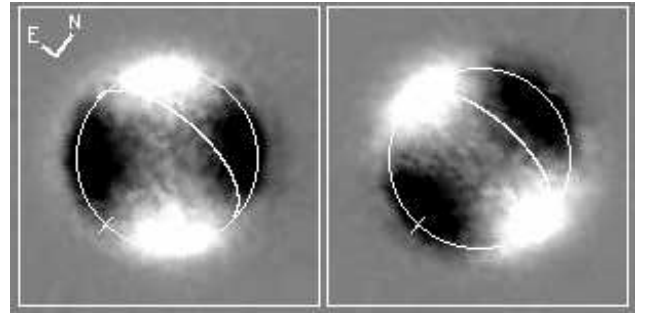}
                 \caption{Stokes $Q$ (left) and $U$ (right) images in the $F775W$ band. The south pole
                 and the equator are indicated. The gray scale is normalized to the central intensity
                 of Stokes $I$ by $\pm 0.02\, I_{\rm center}$. 
                 }
                 \label{stokesqu}    
        \end{figure}

In Fig.~\ref{stokesqu} Stokes $Q$ and $U$ images
in the $F775W$ band are shown. The same highly symmetric quadrant 
pattern is present in all of our data. This butterfly pattern is real and it is practically
impossible to artificially create such a pattern by misalignments of the three
polarization images $i_\ast$ or other spurious effects.
The gray scale is normalized to the central intensity
$I_{\rm center}$ of the planetary disk and goes from $-0.02\,I_{\rm center}$ (black) to $+0.02\,I_{\rm center}$ (white). 
At the center $Q$ and $U$ are essentially zero ($Q/I, U/I \lesssim \pm 0.2~\%$). 

The butterfly polarization pattern is typical for a centro-symmetric scattering geometry. For example, the Rayleigh-scattering 
atmospheres of Uranus and Neptune show this pattern of radial limb polarization \citep{Schmid2006}. 
For all our data the pattern is highly symmetric, indicating that the limb polarization has similar 
strength along the entire limb of Titan. 
The strength of the limb polarization
increases with wavelength until it peaks in the 1 $\mu$m band measurement after which it decreases again.

\begin{table*}
\caption{Polarization results for Titan. The columns give the filter, the central wavelength $\lambda_c$, the
effective wavelength $\lambda_{\rm eff}$, the effective albedo $A_{\rm g,eff}$ for
the photons registered in the broad-band filters, and the \citet{Toon1992} radius $R_{\rm T}$ for $\lambda_{\rm eff}$.
$\langle Q/I\rangle_{\rm m}$, $\langle U/I\rangle_{\rm m}$, $\langle Q_{\rm r}/I\rangle_{\rm m}$, and $\langle U_{\rm r}/I\rangle_{\rm m}$
are the measured disk-integrated polarization parameters.
The parameter $C_{\rm PSF}$ describes the degradation of the polarization measurement due to the PSF smearing effect  (Sect. \ref{s:resolution}), and 
the corrected radial polarization value is given by $\langle Q_{\rm r}/I\rangle = \langle Q_{\rm r}/I\rangle_{\rm m} / C_{\rm PSF}$.
$(Q_{\rm r}/I)_{\rm m}^{\rm max}$ is the measured maximum radial polarization, whereas $(Q_{\rm r}/I)^{\rm max}$ is the modeled value for infinite resolution.
The statistical 1$\sigma$ error for the disk-integrated polarization is estimated to be $\Delta p\leq \pm 0.1\,\%$
and $\Delta C_{\rm PSF} = \pm 0.01$.
In addition, a systematic uncertainty of $(\Delta p)_{\rm syst.}\approx \pm 0.2~\%$ is estimated.
  }
\label{intpol}
\centering
\begin{tabular}{lC{8mm}C{7mm}C{7mm}C{10mm}C{12mm}C{10mm}C{10mm}C{10mm}C{10mm}C{10mm}C{10mm}C{12mm}}
\hline \hline
\multicolumn{5}{c}{}  &\multicolumn{2}{c}{integrated polarization} & \multicolumn{4}{c}{integrated radial polarization}  &  \multicolumn{2}{c}{max. radial pol.} \\
\hline
filter 
& $\lambda_c$ & $\lambda_{\rm eff}$& $A_{\rm g,eff}$    &$R_{\rm T}$
         & $\langle Q/I\rangle_{\rm m}$  
                   & $\langle U/I\rangle_{\rm m}$ 
                              &  $\langle Q_{\rm r}/I\rangle_{\rm m}$    
                                       & $\langle U_{\rm r}/I\rangle_{\rm m}$   
                                             & $C_{\rm PSF}$  & $\langle Q_{\rm r}/I\rangle$
                                                           &$(Q_{\rm r}/I)_{\rm m}^{\rm max}$      & $(Q_{\rm r}/I)^{\rm max}$ \\
& [nm]  & [nm]  &   &  [km] & [\%]   & [\%]   &[\%] &   [\%] &  &  [\%]  &  [\%]   & [\%]  \\ 
\hline
$F250W$      & 273              & 299                 & 0.042     &     2901  &     0.13 &          0.00 &       {\bf 0.99} &      0.03 &         0.84         &      {\bf 1.18}   &  {\bf 1.82}   &   {\bf 4.18}            \\
$F330W$      & 336              & 341                 & 0.051      &    2892  &     -0.81 &        -0.36 &       {\bf 1.10} &     -0.05 &        0.85         &      {\bf 1.30}   &   {\bf 2.37}  &   {\bf 4.40}        \\
$F435W$      & 432              & 448                 & 0.097      &    2871   &      0.06 &         0.08 &        {\bf 1.71} &     -0.09 &        0.86          &      {\bf 1.98}  &   {\bf 3.44}   &   {\bf 5.31}        \\
$F625W$      & 632              & 636                 & 0.219      &    2835   &     0.00 &          0.19 &        {\bf 2.51} &     -0.09 &        0.85          &      {\bf 2.95}  &   {\bf 4.93}  &    {\bf 6.86}      \\
$F775W$      & 768              & 762                 & 0.211      &    2813   &     0.09 &           0.03 &       {\bf 2.97} &     -0.14 &        0.82          &      {\bf 3.63}  &   {\bf 5.19}  &    {\bf 8.70}       \\
$NIC1$         & 1071            & 1018                 & 0.155      &  2771    &      -1.49 &          -1.40 &      {\bf 4.03} &      0.04 &         0.74         &      {\bf 5.45}   &   {\bf 6.90}  &   {\bf 10.49}     \\
$NIC2$        & 2002             & 2022               & 0.094      &    2651     &     -0.30 &         -0.09 &      {\bf 1.99} &      0.00 &          0.58        &      {\bf 3.42}  &   {\bf 3.44}   &    -$^\ast$        \\
\hline
\multicolumn{13}{l}{\vspace{-0.2cm}}\\
\multicolumn{13}{l}{$^\ast$wavelength range of model: 200 nm $< \lambda<$ 1600 nm}\\
\end{tabular}
\end{table*}

We do not see a significant imprint of the north-south asymmetry of Titan (Sect. \ref{s:intensity}) in the polarization images.
To study the polarization along the north-south (and east-west) direction the polarization pattern was aligned with respect to the polar axis of Titan by
\begin{displaymath}
Q_{\rm NS}=Q \cos(2\theta_{\rm NP})-U \sin(2\theta_{\rm NP})\, , 
\end{displaymath}
where $\theta_{\rm NP}$ is the angle between the polarizer reference axis and the polar axis of Titan.

For all filters we calculate disk-integrated Stokes fluxes $\Sigma I$, $\Sigma Q$, and $\Sigma U$ by summing up all 
counts within the integration radius $R_{\rm int}= 0.75"$ from the apparent disk center of the $I$, $Q$, and $U$ images respectively. 
$R_{\rm int}$ includes the halo of the planet which is greater than the nominal limb at $R_{\rm Titan}=0.44"$.
We then calculate disk-integrated fractional polarization parameters for Stokes $Q$ (and similarly $U$):
\begin{equation}
\langle Q/I\rangle_{\rm m} \, ({\rm R_{\rm int}}) = \Sigma Q/\Sigma I \, ;
\label{e:integration}
\end{equation}
$\langle Q/I\rangle_{\rm m}$ and
$\langle U/I \rangle_{\rm m}$ are equivalent to
a measurement with aperture polarimetry, where the aperture is larger 
than the planet. The results for all filters are given in Table \ref{intpol}.
The disk-integrated polarization of Titan is essentially zero ($p < 0.2~\%$) for all filters in agreement with \citet{Veverka1973} and \citet{Zellner1973}, except for $F330W$ and for $NIC1$; these results are not real and can be explained by instrumental effects. 

As explained in Sect.~\ref{NICMOS calibration} $NIC1$ is not well calibrated
for instrumental polarization and between $p_{\rm inst.} \approx 1.2~\%-1.5~\%$ residual polarization is expected.
We are measuring $\langle Q/I\rangle_{\rm m} = 1.5~\%$ despite the expectation that Titan has zero net polarization.
The same instrument offset is also seen at the center of the disk which should be zero because of symmetry reasons.
Similarly for the $F330W$ measurement the center of the Stokes $Q$ and $U$ images is not zero indicating residual instrumental polarization
at the level of $p_{\rm inst.} \approx 0.8~\%$ in $Q$ and $p_{\rm inst.} \approx 0.2~\%$ in $U$ respectively. 

The absence of a net polarization in the disk averages indicates that the limb polarization
has a similar strength along the entire limb for all observed bands.
We note that the disk-integrated
parameters $\langle Q/I\rangle_{\rm m}$ and $\langle U/I \rangle_{\rm m}$
can hardly be affected by inaccuracies in the image centering procedure or other spurious effects due to the data reduction.  
The statistical 1$\sigma$ measuring error is $\Delta p< 0.1~\%$ as estimated for the measuring error of the integrated
radial polarization $\langle U_{\rm r} /I\rangle_{\rm m}$ in Sect.~\ref{s:radial polarization} which is independent of any residual instrumental offset.


\section{The radial polarization}\label{s:radial polarization}

The polarization flux of an object is given by
$p\times I=\sqrt{Q^2+U^2}$. 
However, because of the squares in this formula, 
large systematic bias errors are introduced if the absolute value of one or both measured signals
$|Q|$ and $|U|$ is not significantly higher
than the measuring noise $\Delta Q$ and $\Delta U$.
In our Titan data there is $\Delta Q\approx |Q|$ and $\Delta U\approx |U|$
in the middle of the planetary disk, and between positive and negative quadrants in the butterfly pattern. 
Therefore one should not use the polarized flux $p\times I$ or the normalized polarization $p$ as measuring parameter.
We adopt radial Stokes parameters, which are particularly well-suited for characterizing 
centro-symmetric polarization patterns of planets \citep[e.g.,][]{Schmid2006}.

The radial Stokes parameters $Q_{\rm r}$ and $U_{\rm r}$ 
describe the polarization in radial and tangential direction on the
disk of Titan. They are given by
\begin{eqnarray}\label{e:radpolq}
Q_{\rm r} & = & +Q \cos 2\phi + U\sin 2\phi\,, \\
\label{e:radpolu}
U_{\rm r} & = & -Q \sin 2\phi + U\cos 2\phi\,,
\end{eqnarray}
where $\phi$ is the polar angle of a given position $(x,y)$ on the
apparent planetary disk (disk center $(x_0,y_0)$) with respect
to the polarizer reference direction:
\begin{displaymath}
   \phi = \arctan {x-x_0\over y-y_0}\,.
\end{displaymath}
$Q_{\rm r}>0$ is equivalent to a radial polarization or a polarization
perpendicular to the limb, while $Q_{\rm r}<0$ indicates a tangential
polarization component. $U_{\rm r}$ describes the polarization in the
directions $\pm 45^\circ$ with respect to the radial direction.

\subsection{Radial Stokes $Q_{\rm r}$ and $U_{\rm r}$ images for Titan}\label{s: radial images}

\begin{figure}
        \centering
        \includegraphics[width=4.1cm,angle=90]{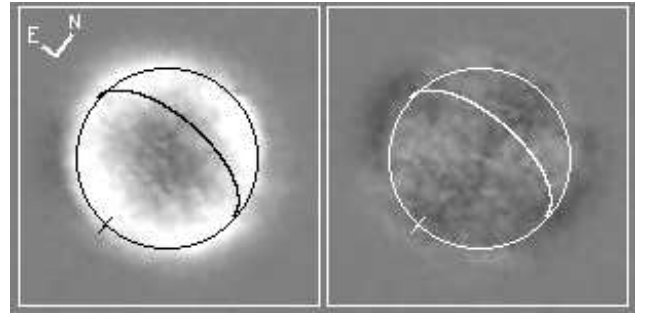}
                 \caption{Radial Stokes $Q_{\rm r}$ (left) and $U_{\rm r}$ (right) images in the $F775W$ band. The south pole
                 and the equator are indicated. The gray scale is normalized to the central intensity
                 of Stokes $I$ by $\pm 0.02\, I_{\rm center}$.}
                 \label{rstokesqu}    
        \end{figure}

\begin{figure}
        \centering
        \includegraphics[width=7.6cm]{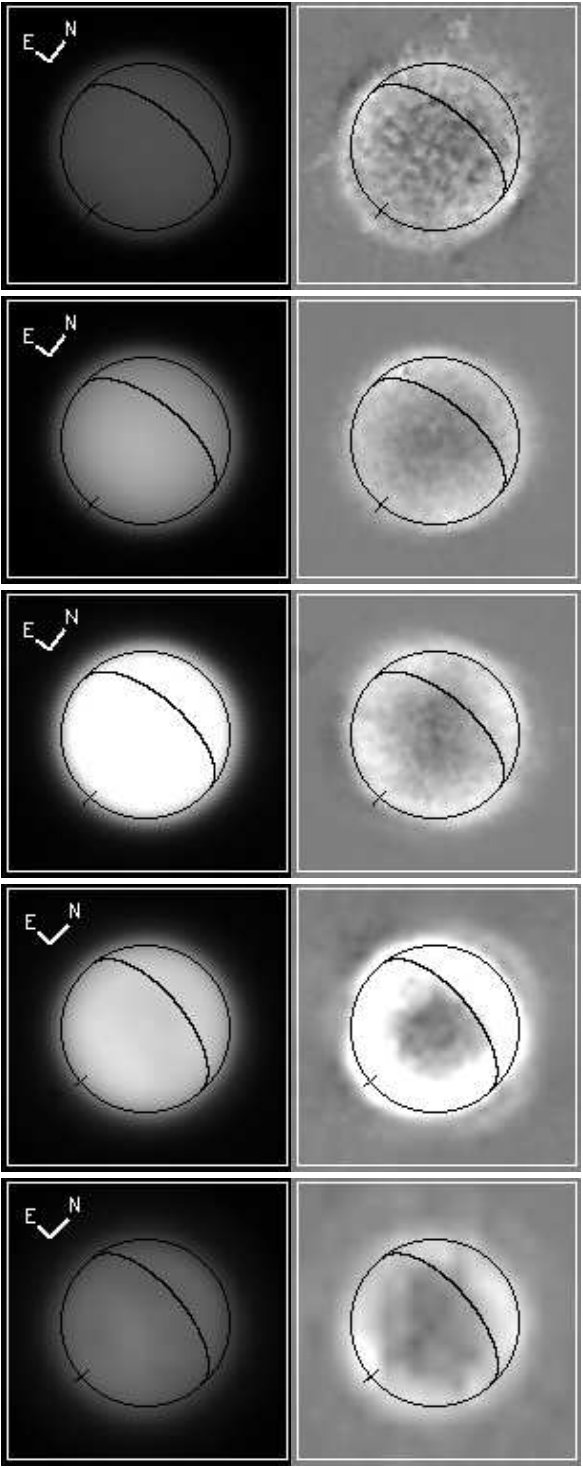}
                 \caption{Intensity (left) and radial polarization images $Q_{\rm r} $ (right) in the $F330W$, $F435W$, $F625W$, $NIC1$, and $NIC2$ band (top to bottom).
                 The intensity images are normalized to $\int_0^R \int_0^{2\pi} Irdrd\phi = A_{\rm g,eff} \cdot \pi R^2$ and the gray scale of the
                 polarization images is scaled to $\pm 2~\%$ of the central intensity $I_{\rm center}$.}
                 \label{rstokesiq}    
        \end{figure}

For all observed bands we calculated radial polarization images $Q_{\rm r}$ and $U_{\rm r}$. 
In case of the $F330W$ and $NIC1$ observations the Stokes $Q$ and $U$ images were first 
corrected for residual instrumental polarization (see Sect. \ref{sectqu}), derived from the disk center which is expected to be unpolarized.
Figure \ref{rstokesqu} shows $Q_{\rm r}$ and $U_{\rm r}$ for Titan in the 
$F775W$ filter. 
In all of our data the limb polarization is
clearly visible as a bright ring with positive $Q_{\rm r} $ polarization and essentially zero $U_{\rm r} $ polarization.
Except for the $F250W$ filter, the level of the $U_{\rm r} $ polarization is typically about 10 times lower than the positive $Q_{\rm r} $
signal along the limb. For the $F250W$ filter the $U_{\rm r} $ polarization is about 3 times lower than $Q_{\rm r} $.
Thus $Q_{\rm r} $ dominates in all filters.
In Figure \ref{rstokesiq} we show images of $Q_{\rm r} $ as well as corresponding Stokes $I$ for the rest of our data.
The gray scale of the radial polarization images in Figs. \ref{rstokesqu} and \ref{rstokesiq}
is scaled to the central intensity $I_{\rm center}$ by $\pm 0.02 \cdot I_{\rm center}$, and the intensity
images are normalized to the effective albedo $A_{\rm g,eff}$ according to Eq. (\ref{inorm}).

\subsection{Polarization as function of radius}\label{s: function of radius}

      \begin{figure}[t!]
        \centering
        \includegraphics[width=8.2 cm]{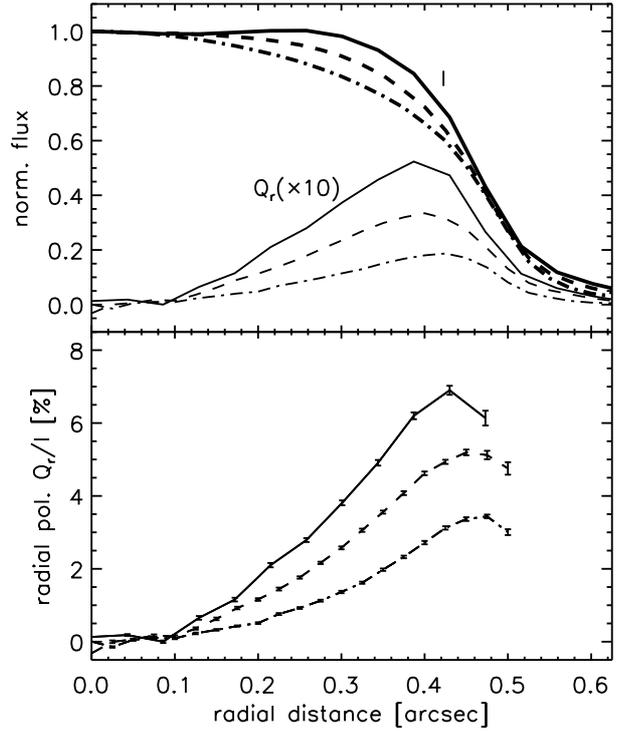}
                 \caption{Titan radial profiles for intensity $I$ (upper panel, thick lines) and radial
 		 polarization $Q_{\rm r} $ (upper panel, thin lines) in the $F435W$ (dash-dot), $F775W$ (dash) and $NIC1$ (solid) filters. Both $I$
  		and $Q_{\rm r} $ are normalized to the peak flux at $r=0$. The lower panel 
 		 shows the corresponding normalized radial polarization
		  $Q_{\rm r} /I$.}
                 \label{rprofiles}    
        \end{figure}

The observed polarization of Titan
is essentially centro-symmetric, and contrary to the strong north-south albedo asymmetry, the
corresponding imprint in the radial polarization flux is either absent or much weaker. In the fractional radial polarization images $Q_{\rm r}/I$ 
we see marginal north-south differences with higher polarization in the north for the bands shorter than $1~\mu$m, the opposite effect 
in the $NIC1$ band, and about equal polarization for the $NIC2$ band.
However, these results are not significant and on the order of our systematic uncertainties $\Delta p/p\approx 0.05-0.1$.

In the infrared there is a similar weak indication for an east-west asymmetry of $Q_{\rm r}/I$, whereas for the $NIC1$ band the
polarization seems to be higher in the east, which is inverted for the $NIC2$ band. 
It also seems that for wavelengths shorter than $1~\mu$m the maximum polarization is 
higher at the eastern and western limbs than in the north and south, whereas for the $NIC1$ band the polarization in east-west and north-south is of about equal strength, and
for $NIC2$ the polarization is higher at the poles.
However, the quality of our data is not good enough to draw firm conclusions, and at least part of the polarization differences mentioned above 
could be due to systematics in the data reduction, 
e.g., slight misalignments of corresponding polarization images coupled with strong intensity gradients at the limb.

The detection of structure in the limb polarization would be very interesting for investigating local haze properties, 
e.g., such as particle size differences of the photochemical haze between the morning and the evening limb of Titan.
Observations with higher polarimetric sensitivity and higher spatial resolution 
are required for such studies.

For the moment, assuming a rotational symmetry for the polarization structure seems to be
a reasonable first approximation,
and we construct rotationally averaged, radial
profiles for the polarization, the normalized polarization, and the 
intensity. The results for the
$F435W$, $F775W$, and $NIC2$ filters are shown in Fig. \ref{rprofiles}.

In all filters the radial profiles look very similar.
The polarization $Q_{\rm r} $ in the disk center at 
$R=0$ is essentially zero. The normalized radial polarization $Q_{\rm r} /I$ 
increases steadily with radius until it peaks at around $R_{\rm Titan}=0.44"$ 
after which it decreases again until the photon noise starts dominating
the measurements. 
Similarly, the radial polarization flux $Q_{\rm r}$ also increases with radius but only up to a radius slightly smaller than $R_{\rm Titan}$,
and then it decreases farther out to zero in step with the intensity profile. 

Both $Q_{\rm r} /I$ and $Q_{\rm r}$ increase with wavelength until they peak
in the $NIC1$ band at 1 $\mu$m. Then, in the $NIC2$ band at 2 $\mu$m the polarization
has again significantly dropped.
Using tentative fits for the radial profiles, we measure maximum 
fractional radial polarization values $Q_{\rm r} /I$ of $1.8~\%$ for the $F250W$ band, up to $6.9~\%$ for the $NIC1$ band, 
and $3.4~\%$ for the $NIC2$ band. The values for all filters are given in Table \ref{intpol}. 
However we note that these values are not yet corrected for the PSF smearing effect described in Sect. \ref{s:resolution}.

\subsection{Disk-integrated radial polarization}\label{s: disk-integrated radial polarization}

Similar to the calculation of disk-integrated Stokes parameters 
in Sect. \ref{sectqu}
disk-integrated radial Stokes parameters $\langle Q_{\rm r} /I\rangle_{\rm m}$ and $\langle U_{\rm r} /I\rangle_{\rm m}$ 
are calculated for all filters. 
Because of the intrinsic rotational symmetry of Eqs. (\ref{e:integration}), (\ref{e:radpolq}), and (\ref{e:radpolu}),  
$\langle Q_{\rm r} /I\rangle_{\rm m}$ and $\langle U_{\rm r} /I\rangle_{\rm m}$ have
the additional advantage that any instrumental polarization offset or gradient cancels out.
Therefore, we do not need to correct
the instrumental offset described in Sect. \ref{sectqu} for the $F330W$ and $NIC1$ filter to calculate corresponding disk-integrated
radial polarization values. This was also verified by calculating $\langle Q_{\rm r} /I\rangle_{\rm m}$ and $\langle U_{\rm r} /I\rangle_{\rm m}$ 
for the $NIC1$ data, both with and without correction of the instrumental offset, which showed that the difference is less than $\Delta p = 0.01~\%$. 

        \begin{figure}
        \centering
        \includegraphics[width=8.2cm]{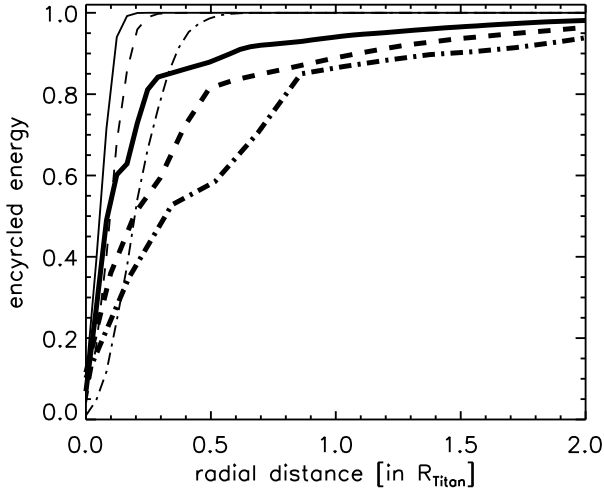}
                 \caption{Encircled energies for HST PSFs (thick lines) and Gaussian PSFs (thin lines): ACS $F625W$ (solid), 
                 NIC1 $POL0S$ (dashed), and NIC2 $POL0L$ (dash-dot).}
                 \label{encenergy}    
        \end{figure}

  \begin{figure}
        \centering
        \includegraphics[width=8.2cm]{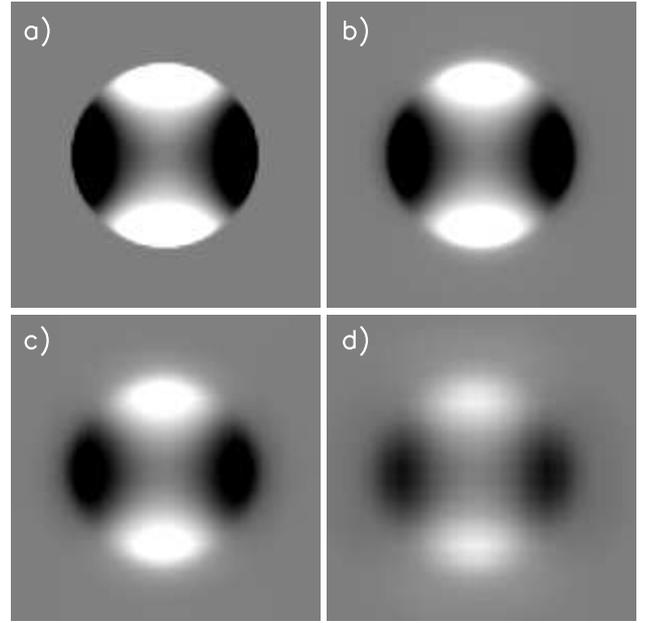}
                 \caption{Modeling of the degradation of the Titan polarization $Q$ 
		due to the cancelation of opposite polarization components $+Q$ and $-Q$
		caused by the limited resolution;
		(a) unlimited resolution, (b) HST PSF at 630~nm ($F625W$), 
		(c) HST PSF at 1$\mu$m ($NIC1$), and (d) HST PSF at 2$\mu$m ($NIC2$). 
  		The gray scale spans for all panels the range 
  		from $-2~\%$ (black) to $+2~\%$ (white) of the peak intensity 
  		of the initial perfect intensity image.}
                 \label{smearplot}    
        \end{figure}

A strong positive signal is obtained for the disk-integrated radial
polarization $\langle Q_{\rm r} /I\rangle_{\rm m}$,
while $\langle U_{\rm r} /I\rangle_{\rm m}$ is essentially zero (Table \ref{intpol}). 
The 1$\sigma$ measuring error is $\Delta p <0.1~\%$
as estimated from the assumption $\langle U/I \rangle_{\rm m} = 0$. 
Both $\langle Q_{\rm r} /I\rangle_{\rm m}$ and  $\langle U_{\rm r} /I\rangle_{\rm m}$
are only very weakly affected by small asymmetries caused by inaccuracies in the image centering procedure or other spurious effects due to the data reduction.
However, strong asymmetric perturbations such as strong ghosts could probably bias the result. Anyway, except for $NIC1$ we do not see any ghosts in our
$Q_{\rm r}$ and $U_{\rm r}$ images (Sect. \ref{s: radial images}). In case for $NIC1$ it is known that weak ghosting is present for bright sources but we estimate that 
the impact on the integrated radial polarization is less than $\Delta p /p = 0.01$ (see also Sect. \ref{s: results}).

The integrated radial polarization $\langle Q_{\rm r} /I\rangle$ is a good 
parameter for characterizing the overall limb polarization of a planet. Since 
$Q_{\rm r} $ is either positive or close to zero everywhere on the disk
no polarization compensation effect is present. Furthermore, there is
$\langle Q_{\rm r} /I\rangle\gg \langle U_{\rm r} /I \rangle\approx 0 $, 
so that we can approximate
\begin{displaymath}
    \langle p_{\rm r} \rangle= \sqrt{\langle Q_{\rm r} /I\rangle^2+\langle
      U_{\rm r} /I\rangle^2} \approx \langle Q_{\rm r} /I \rangle \,.
\end{displaymath}

\subsection{Correction for the PSF smearing effect} \label{s:resolution}
   
The point spread functions of HST ACS and NICMOS have a finite width given by the telescope diffraction $\sim$$\lambda/D$, and
they are affected by optical aberrations,
geometric distortions, and in case of ACS a long-wavelength halo produced by the detector itself. 
This leads to extended PSF wings which limit the resolution of the HST observations. 
Therefore, the measured integrated limb polarizations $\langle Q_{\rm r} /I\rangle_{\rm m}$ and $\langle U_{\rm r} /I\rangle_{\rm m}$ 
need to be corrected by the inverse of a degradation factor $C_{\rm PSF}$, accounting for a polarization
cancelation due to PSF smearing. 

We produced simulated ACS and NICMOS PSFs for all wavelengths, using the  
Tiny Tim\footnote{http://tinytim.stsci.edu} PSF 
simulation software package for the HST \citep[see][]{Krist2011}.
Figure \ref{encenergy} compares the encircled energy of these PSFs to the encircled energy of Gaussian PSFs with ${\rm FWHM}=\lambda/D$.
One can see that for the Gaussian PSFs essentially all the energy is 
contained within half the radius of Titan. However, in the $NIC2$ band up to $45~\%$ of the energy is smeared 
over an area larger than half of the radius of Titan. For the ACS filters and $NIC1$ the fraction
originating from $R>0.5~ R_{\rm Titan}$ is considerably smaller but still about $15~\%$ and $20~\%$
respectively.

Because of these extended PSF wings, the opposite polarization components 
$+Q$ and $-Q$ overlap and cause a reduction in the resulting net
polarization. In the most extreme case of an unresolved 
centro-symmetric planetary disk, the polarization cancelation would be perfect 
and only a zero net polarization level could be measured. 
The compensation effect
is stronger for longer wavelengths, where the diffraction limited
spatial resolution of HST is not as good, and at the same time the PSF wings are stronger.

For an estimate of the polarization cancelation, we adopted our
haze scattering model for the expected polarization pattern (see Sect. \ref{modeling}).
From the model we constructed two-dimensional intensity 
images for $i_0$, $i_{90}$, $i_{45}$, and $i_{135}$, from which the corresponding
Stokes $Q$ and $U$ images can be 
calculated by $Q=i_0 - i_{90}$ and $U=i_{45} - i_{135}$.
Similarly, we constructed smeared Stokes images $Q_{\rm s}$ and 
$U_{\rm s}$ by folding the $i_\ast$ images with the simulated HST PSFs.
Figure~\ref{smearplot} illustrates the cancelation effect in the
Stokes $Q$ image due to the HST PSF of ACS at $F625W$, $NIC1$ at 1$\mu$m, and
$NIC2$ at 2$\mu$m. 

The $Q$ and $U$ images can then be converted into radial
polarization images $Q_{\rm r} $ and $U_{\rm r} $, as for the observations. 
From $Q_{\rm r} $ the integrated radial 
polarizations $\langle Q_{\rm r} /I\rangle$ and $\langle Q_{\rm r} /I\rangle_{\rm s}$ are calculated for 
the different filters in the same way as for the observations. The ratio between the clean 
and the smeared polarization then yields the factor for the expected degradation of the disk-integrated
radial polarization 
\begin{equation}
C_{\rm PSF}=\frac{\langle Q_{\rm r} /I\rangle_{\rm s}}{\langle Q_{\rm r} /I\rangle}\,.
\end{equation}
The corresponding
values of $C_{\rm PSF}$ and the corrected limb polarization $\langle Q_{\rm r}/I\rangle$ for all filters are given in Table \ref{intpol}. 
Especially for $NIC2$ the degradation factor $C_{\rm PSF} =0.58$
is low, while for the ACS bands and for $NIC1$ the degradation factor is about $C_{\rm PSF}=0.85$ and $C_{\rm PSF}=0.74$ respectively.
The statistical 1$\sigma$ error of the degradation factor is estimated to be about $\Delta C_{\rm PSF} = \pm 0.01$.
 
The polarization cancelation depends not on 
the strength, but on the geometric structure of the polarization pattern. 
Rayleigh scattering models indicate that the polarization
pattern is very similar for the different model parameters \citep[see][]{Schmid2006}. Thus the
degradation depends not significantly on the exact haze scattering parameters of the planet. 
A sanity check by recalculating $C_{\rm PSF}$ using a Rayleigh scattering model indeed shows that
within our 1 $\sigma$ error bars the results for $C_{\rm PSF}$ were identical.


\section{Comparison with limb polarization models}\label{modelsect}

Titan is an excellent test case for detailed studies of the scattering polarization from a hazy atmosphere,
and accurate scattering and polarization parameters are available from the in situ measurements of the Huygens landing probe
\citep[e.g.,][]{Brown2010, Tomasko2008}.
In the next section we describe the basic atmospheric structure that we used for
our radiative transfer model, which is described in Sect. \ref{modeling}, and in Sect. \ref{s: results} we compare the model with our limb polarization measurements and literature values for the geometric albedo $A_{\rm g}$ and the quadrature polarization $p(90^\circ)$ of Titan.

\subsection{Atmospheric parameters}

For our model we assume an atmosphere ranging from \mbox{0-1300 km}. This includes
Titan's troposphere with its well defined tropopause at $\sim$44 km (112 mbar), 
the stratosphere with the stratopause located at $\sim$260-310 km (0.22-0.08 mbar; \citealt{Fulchignoni2005}, \citealt{Vinatier2007}), 
the mesosphere with the mesopause at $\sim$494 km (0.002 mbar; \citealt{Fulchignoni2005}), and the thermosphere ranging up to $\sim$1300 km ($1.4\cdot 10^{-8}$ mbar).

The atmospheric composition is predominantly N$_2$, with CH$_4$ and H$_2$ the second and third most abundant molecules respectively. 
Near the surface CH$_4$ has an abundance of $\sim$5~\%, falling to $\sim$1.4~\% in the stratosphere
\citep{Brown2010}. Because of its long chemical lifetime H$_2$ is essentially
uniformly mixed throughout the atmosphere with a mixing ratio $\sim$0.1~\% \citep{Courtin2008}. 
A thick haze layer is located in the stratosphere, and a second detached haze layer lies just above the mesopause at around 500 km 
altitude \citep{Porco2005}.
Above 700-800 km most of the methane is destroyed by photolysis.

      \begin{figure}
        \centering
        \includegraphics[width=8.8cm]{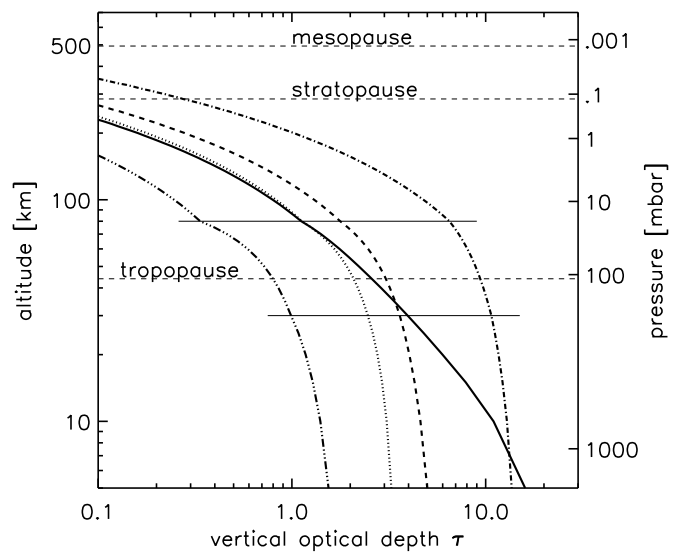}
                 \caption{Vertical optical depth $\tau$ for $\lambda=445$ nm (dash-dot), $\lambda = 775$ nm (dash), $\lambda = 940$ nm (dot), 
                 $\lambda = 1000$ nm (methane absorption band, solid),
                 and for $\lambda =1580$ nm.
                }
                 \label{tau}    
        \end{figure}

For simplification, we assume the atmospheric composition to be solely made out of methane and nitrogen.
The methane mole fraction $f_{\rm CH_4} (h)$ between $h=0-144$ km altitude was taken from \citet{Brown2010}, whereas 
above 140 km $f_{\rm CH_4}(h)$ was assumed to linearly drop to zero at an altitude of 600 km. 
We use the temperature $T(h)$ and density $\rho(h)$ profiles provided by the Huygens Atmospheric Structure Instrument 
(HASI)\footnote{http://atmos.nmsu.edu/PDS/data/hphasi\_0001/DATA/PROFILES/}, and
the total molecular number densities and the Rayleigh scattering optical depth $\tau_{\rm Ray}(h, \lambda)$ are then calculated from the ideal gas law 
and the gas column density (see appendix).   

For the methane absorption coefficients $\kappa(T, \lambda)$ we use the formula given by \citet{Karkoschka2010}. We note that below
$\lambda=1~\mu$m these coefficients are generally close to those by \citet{Karkoschka1998}.

A detailed model of the aerosol properties of Titan is given by \citet{Tomasko2008}, based on measurements of the DISR instrument on the Huygens landing
probe \citep{Tomasko2002}. 
The haze optical depth per unit path length 
 $\tau_{\rm haze}(h, \lambda)$ is derived for three altitude regions, i.e., above 80 km, between 30-80 km, and below 30 km. 
On the one hand, \citet{Tomasko2008} describe the wavelength dependence by three different power laws, corresponding to the three altitude regions.
Then on the other hand, the cumulative optical depth increases with decreasing altitude. Between 0-30 km and 30-80 km the increase is linear but with two different
slopes, and above 80 km the increase is exponential with a scale height of 65 km. 

The vertical optical depth is shown in Fig.~\ref{tau} for 
five different wavelengths between $\lambda = 445$ nm and $\lambda = 1580$ nm, including the methane absorption band at $\lambda = 1~\mu$m. 
For wavelengths $\lambda < 1~\mu$m most of the light is scattered in the 
stratosphere between $\sim$100-300 km altitude where $\tau\approx 1$, whereas for $\lambda=1.6~\mu$m the light penetrates down to about 30 km altitude.  
The impact of the methane absorption is particularly strong in the troposphere but has almost no effect for higher altitudes. 
 
Similar to $\tau$, the single scattering albedo $\omega_{\rm haze}(h, \lambda)$ can be split into wavelength and altitude dependent parts, whereas
typically $\omega_{\rm haze}(h, \lambda) \approx 0.8-1$.
We adopt the altitude model of \citet{Tomasko2008}.
Between 0-30 km, 30-80 km, and above 144 km respectively, we assume $\omega_{\rm haze}(h, \lambda)$ to be constant with altitude, using the values given in Table 2 of \citet{Tomasko2008}, whereas between 80-144 km we linearly interpolate between the adjacent regions.
For the region between 80 km to 144 km \citet{Tomasko2008}
suggest that new material is incorporated in the aerosols as they fall, and that the aerosols grow in size. In the few kilometers above the surface
they also see some weak evidence of a decrease of $\omega_{\rm haze}(h, \lambda)$, i.e., reversing the general trend with altitude.
The wavelength dependence of $\omega_{\rm haze}$ for the region above 144 km and 30-80 km is given in Figure 48 of \citet{Tomasko2008}, and the same dependence is assumed for the region 80-144 km.  For the region below 30 km
we adopt a three-dimensional polynomial fit to the values given in Table 2 of \citet{Tomasko2008}.

For the surface we assume a diffusely scattering surface with constant albedo $A_{\rm s}=0.2$ for all wavelengths.
This is a strong simplification and according to the literature the surface albedo of Titan varies between $A_{\rm s}=0.1-0.3$, depending on wavelength and also the season
of Titan \citep[e.g.,][]{McKay1989, Tomasko1997, Tomasko2008}. Anyway, we tested different models using a range of \mbox{$A_{\rm s}=0.1-0.3$}, and 
the variation from $p(0^\circ)|_{A_{\rm s}=0.2}$ is less than $\Delta p(0^\circ)/p(0^\circ) \approx 0.1$. 

The wavelength range of our model is restricted by the \citet{Tomasko2008} values for $\tau_{\rm haze}(h,\lambda)$ and $\omega_{\rm haze}(h,\lambda)$, 
which are only given for a wavelength range of \mbox{400-1600 nm}.
On the one hand, we extended this range towards the UV by extrapolating the parameters down to 200 nm. To first order this is valid because between 200-400 nm 
we do not expect strong spectral or altitudinal features in $\tau_{\rm haze}(h,\lambda)$ and $\omega_{\rm haze}(h,\lambda)$. 
On the other hand, we did not extrapolate the parameter range towards the red end because above $\lambda \approx 1~\mu$m the spectral variation in 
$\omega_{\rm haze}$ is more complex, and because the extrapolation interval to include the $NIC2$ band at $2.2~\mu$m was too large. 
Therefore, the final spectral range of the model covers 200-\mbox{1600 nm}, which includes all our polarization data except the $NIC2$ polarimetry.

\subsection{Radiative transfer code}\label{modeling}

We use an extended version of the Monte Carlo scattering code described by \citet{Buenzli2009}. 
Basically, the code calculates random walk histories of many photons entering the atmosphere, and follows their direction and polarization change
until they are absorbed or they escape. 
The intensity and polarization spectra of the planet can then be established for different lines of sight, 
and in the case for backscattering ($\alpha=0^\circ$) as a function of radial distance from the disk center.
For the calculation, the spherical model atmosphere is assumed to be rotationally homogeneous, 
consisting of multiple locally plane parallel layers. 
The incident radiation is a parallel beam of unpolarized photons, whereas despite multiple scattering, 
the photons emerge at the same point where they entered into the atmosphere. 

The scattering processes are described by probability density functions, derived from the appropriate phase matrices of the scattering particles \citep[see also][]{Schmid1992}.
For scattering on haze particles, the code allows for scattering matrices of the form
\begin{equation}\label{scattering matrix}
\mathbf{F} (\theta) = \left(
           \begin{matrix} 
            F_{11}(\theta) & F_{12}(\theta)   & 0 & 0 \\
            F_{12}(\theta) & F_{11}(\theta)    & 0 & 0 \\
            0 & 0 & F_{33}(\theta) & 0 \\
            0 & 0 & 0 & 0   \\
           \end{matrix}\right) \, ,
\end{equation}
where $\theta$ is the scattering angle. The code does not consider circular polarization because these effects are expected to be very small and negligible
for a simple scattering model.

   \begin{figure}
        \centering
        \includegraphics[width=7cm]{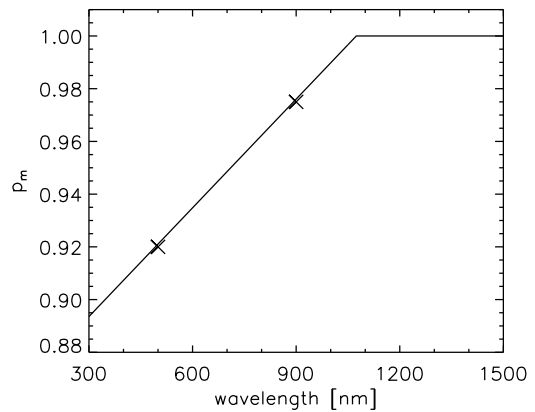}
                 \caption{Wavelength dependence of the single scattering polarization parameter $p_m$ defined by Eq. (\ref{pm}). The fit values to the
                 \citet{Tomasko2008} curves for the blue and the red channel are also indicated. 
                 }
                 \label{pm dependence}         
        \end{figure}

We used the aerosol scattering phase functions $F_{11}(\theta)$
given in tabulated form by \citet{Tomasko2008} for two altitude regions, above 80 km and below 80 km, and for different 
wavelengths ranging between 355 nm to 5166 nm. Typically a phase function is given about every 100 nm for $\lambda < 1$ $\mu$m and about every 200 nm for 
$1~\mu{\rm m} < \lambda < 1.5~\mu{\rm m}$, and for our model calculations we linearly interpolate between these values. Anyway, it turned out that for the polarization results the dependence on $\lambda$ is very weak. This was tested by running different model implementations using only the tabulated phase functions with the closest match to $\lambda$, and within our observational error bars the calculated limb polarization was the same.
 
\citet{Tomasko2008} also derive a single scattering polarization fraction ($-F_{12}/F_{11}$) for the blue ($470-530$ nm) and the red ($880-970$ nm).
To first order, we find a very tight fit to their model using a scaled Rayleigh-like single scattering polarization dependence according to
\begin{equation}\label{pm}
\frac{F_{12}(\theta)}{F_{11}(\theta)} = p_m \frac{\cos^2(\theta) - 1}{\cos^2(\theta) + 1}\, ,
\end{equation}   
with $p_m=0.920$ for the blue channel and $p_m=0.975$ for the red channel.
For our model we use $p_m (\lambda)$ with
a linear slope, going through the values of the red and blue channel, and $p_m (\lambda) = 1$
above $\lambda \approx 1~\mu$m as shown in Fig. \ref{pm dependence}.

For $F_{33}(\theta)$ we use the same dependence as for Rayleigh scattering according to
\begin{equation}
\frac{F_{33}(\theta)}{F_{11}(\theta)} = \frac{2\, {\cos(\theta)}}{\cos^2(\theta) + 1} \, .
\end{equation}

The atmospheric parameters used for our model are described in the previous section. 
Our calculations include Rayleigh scattering, aerosol scattering, and methane absorption but we neglect Raman scattering, which has only a very small effect on 
the reflectivity and polarization \citep[e.g.,][]{Sromovsky2005}.
The model atmosphere consists of 47 different layers above a diffusely scattering surface, and the 
models are run with
$10^9 - 10^{10}$ photons depending on wavelength, such that the statistical error of the fractional polarization is $\Delta p \leq \pm 0.1~\%$.

\subsection{Results}\label{s: results}

\begin{figure}[t!]
        \centering
        \includegraphics[width=9cm]{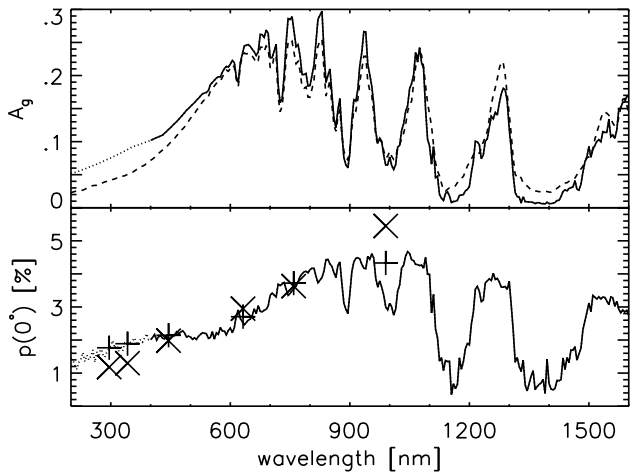}
                 \caption{Comparison of our model results for the limb polarization with literature values. \textit{Top panel:} geometric albedo $A_{\rm g}$ from the literature 
                 (dashed, see also Fig. \ref{figalbedo}) and from our model (solid).
                 \textit{Bottom panel:} integrated limb polarization $p(0^\circ)$ of our model for the complete wavelength range (solid), and the HST filter pass bands ($+$), as well
                 as our HST polarimetry observations ($\times$) given in Table \ref{intpol}.  
                 }
                 \label{fig: results limbpol}         
        \end{figure}

Our model calculates the geometric albedo $A_{\rm g}$, the quadrature polarization $p(90^\circ) = \langle Q/I\rangle(90^\circ)$, and the 
limb polarization $p(0^\circ)=\langle Q_{\rm r} /I\rangle$. 
Fig. \ref{fig: results limbpol} compares in the top panel the calculated geometric albedo with observational data (see Sect. \ref{observations}).
We find a good qualitative agreement for the complete wavelength range. 
Above 500 nm the agreement
outside strong absorption bands is better than $\Delta A_{\rm g}/A_{\rm g} = 0.2$, whereas for the strong absorption around $1.15~\mu$m and $1.45~\mu$m the agreement 
is $\Delta A_{\rm g}/A_{\rm g} \approx 0.3$.
Below 500 nm the agreement gets worse with decreasing wavelength. At 400 nm it is $\Delta A_{\rm g}/A_{\rm g} \approx 0.4$ and at 300 nm it is only
$\Delta A_{\rm g}/A_{\rm g} \approx 0.75$.
We note that the disagreement below
400 nm is not alarming because there the albedo is very low. Furthermore, it could also origin from our questionable extrapolation of the \citet{Tomasko2008} haze parameters from 400 nm to 200 nm.

\begin{figure}[t!]
        \centering
        \includegraphics[width=9cm]{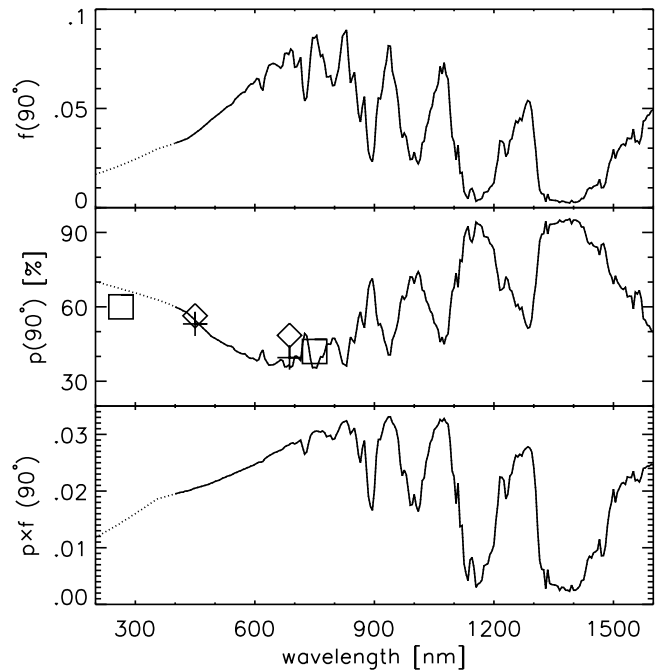}
                 \caption{Comparison of our model results at quadrature phase  $\alpha=90^\circ$ with literature values.
                  \textit{Top panel:} Reflectivity $f(90^\circ)$.
                  \textit{Middle panel:} integrated quadrature polarization $p(90^\circ)$ of our model for the complete wavelength range (solid), and the results from the Pioneer 11 
                  ($\diamond$)
                 and Voyager 2 ($\square$) spacecrafts. The model results in the Pioneer 11 pass bands are also indicated ($+$).
                 \textit{Bottom panel:} integrated polarization flux $p\times f~ (90^\circ)$. 
                 }
                 \label{fig: results quadrature}         
        \end{figure}

The limb polarization results for the model and our measurements are shown in the bottom panel of Fig. \ref{fig: results limbpol}.
Between $400-900$ nm the agreement is good with $\Delta p/p \approx 0.1$ for the $F625W$ pass band, and $\Delta p/p \le 0.05$ for the $F435W$ and F775 pass bands. 
Below 400 nm the model seems to systematically overestimate the polarization, yielding a discrepancy of $\Delta p/p \approx 0.3$. 
Qualitatively this UV-offset agrees with our result for the geometric albedo $A_{\rm g}$, which also seems to be systematically too high in the UV.
Above 900 nm our $NIC1$ polarization is much higher than the model result and $\Delta p/p \approx 0.25$. We could not conclusively determine whether this discrepancy
is caused by an issue of the measurement, the modeling, or both. 

It is known that the polarimetric calibration of $NIC1$ has some deficiencies such as residual instrumental polarization of $p_{\rm inst.}\sim$1.5~\% and weak ghosting 
(see Sect. \ref{NICMOS calibration}). However, because of the intrinsic symmetry of the radial Stokes parameters $Q_{\rm r}$ and $U_{\rm r}$ the residual instrument polarization has no 
effect on $\langle Q_{\rm r} /I\rangle$ (see Sect. \ref{s: disk-integrated radial polarization}). The ghosting on the other hand is asymmetric and thus could have an impact on 
$\langle Q_{\rm r} /I\rangle$. However, in our data we only see very weak ghosts which we do not believe to have an effect at the percent level. 
Finally, it could also be that the real PSF is different from the adopted PSF used for the efficiency correction 
$C_{\rm PSF}$ (Sect. \ref{s:resolution}). An overcompensation of the PSF smearing could explain a $\Delta p/p$ of a few percent but
certainly not the full discrepancy between model and data.

Concerning the model parameters for the $NIC1$ wavelength range and their impact on $\langle Q_{\rm r} /I\rangle$,
we tested the effect of the surface albedo $A_{\rm s}$, 
the single scattering polarization parameter $p_m$, and the methane fraction as functions of altitude. 
Using a range of \mbox{$A_{\rm s}=0.1-0.3$} showed that the impact of $A_{\rm s}$ is less than $\Delta p / p \approx 0.1$.
Similarly, setting $p_m=1$ everywhere cannot explain the discrepancy and the effect is less than $\Delta p / p \approx 0.05$. 
Finally, we set the methane mole fraction $f_{\rm CH_4}=0$ for $h>80$ km to check the impact of the methane absorption in the
stratosphere and mesosphere. This basically gets rid of the absorption dips in $\langle Q_{\rm r} /I\rangle$ but only has a minor impact outside the 
absorption bands. Since $\langle Q_{\rm r} /I\rangle$ is flux weighted, we conclude that the overall impact of the methane absorption on the limb polarization
is much less than $\Delta p / p \approx 0.1$. Therefore, the cause of the disagreement between the model and the $NIC1$ measurement remains uncertain. 

Titan full disk phase curves for intensity and polarization have been obtained by the Pioneer 11 \citep{Tomasko1982} and the Voyager 2 \citep{West1983} spacecrafts. 
The Pioneer 11 data were obtained in the $B$ and $R$ bands, covering phase angles between $\alpha = 28^\circ - 96^\circ$, whereas the Voyager 2 data were taken in 
the near UV at 264 nm and the near IR at 750 nm, covering phase angles $\alpha = 2.7^\circ - 154^\circ$.
The second panel of Fig. \ref{fig: results quadrature} compares our model with the quadrature polarization at $\alpha=90^\circ$ measured by Pioneer 11 and Voyager 2. 
For three bands we find a good agreement better than $\Delta p / p \approx 0.1$, whereas for the Pioneer 11 $R$ band the agreement is not as tight but still 
at the level $\Delta p / p \approx 0.25$.

We note that the quadrature polarization $p(90^\circ)$ is generally higher in the absorption bands because
multiple scattering is strongly reduced, and
the reflection is dominated by strongly polarized single scattering at $\sim$$90^\circ$. For the limb polarization $p(0^\circ)$ this $A_{\rm g}-p$ correlation is reversed.
Multiple scattering and a low single scattering albedo $\omega$ are required for producing a strong polarization signal since the total polarization is mainly produced by second 
order but also higher order scatterings.


\section{Discussion and conclusions}\label{conclusions}

We present disk resolved imaging polarimetry of Titan, measured with the HST for the UV to the near-IR spectral region. 
From these observations, we derive the disk-integrated radial limb polarization $\langle Q_{\rm r} /I\rangle$ (Table \ref{intpol}) for various filter pass bands,
and we compare our results with the polarization of a model atmosphere. For the model we use a radiative transfer code presented in \citet{Buenzli2009}, adopting Titan 
atmosphere parameters from the literature, which were mainly derived from the Huygens landing probe \citep[e.g.,][]{Tomasko2008}. 

The geometric albedo $A_{\rm g}$ and the quadrature polarization $p(90^\circ)$ are important reference quantities for
characterizing the scattering properties of a reflecting atmosphere. 
Therefore, we derive the reflected-flux weighted geometric albedo
$A_{\rm g,eff}$ for the used filter pass bands (Table \ref{intpol}), using the spectrophotometry of \citet{McGrath1998}, \citet{Karkoschka1998}, and \citet{Negrao2006}; and we
compare our model results for $p(90^\circ)$ with measurements obtained by the Pioneer 11 \citep{Tomasko1982} and the Voyager 2 \citep{West1983} spacecrafts.

A comparison between the model and our observations for $\langle Q_{\rm r} /I\rangle (\lambda)$, as well as a comparison between our model and literature values 
for the geometric albedo $A_{\rm g} (\lambda)$ and the quadrature polarization $p(90^\circ, \lambda)$ are given in Fig.~\ref{fig: results limbpol} and Fig.~\ref{fig: results quadrature}.

\subsection{Detection of the limb polarization.}

For all observed filter bands, our data show a strong limb polarization of several percent, 
as expected from previous polarization measurements of Titan taken at
larger phase angles.   
To our knowledge, this is the first time that the limb polarization of Titan has been measured, and
we did not find any previous earth-bound imaging polarimetry which resolved Titan.

Within the resolution limits of the observations, the measured limb polarization for Titan is centro-symmetric. 
This is similar to observations of Uranus and Neptune \citep{Schmid2006} but for Titan the polarization is much stronger. 
On the other hand, similar polarization strength can be found in observations of Jupiter but there the limb polarization is essentially
only present at the poles \citep[e.g.,][]{Schmid2011}, indicating thick polar haze layers and non-polarizing reflection from the clouds along the equator. 

The polarimetric sensitivity and the resolution of our data are not sufficient to detect variations of the polarization signal along the limb of Titan, 
as it is seen in albedo observations. In Sect.~\ref{s: function of radius} we pointed to some tentative limb polarization structure, which could be present in our data.
Additional observations of Titan with increased polarimetric sensitivity and resolution are required to progress in this direction.

Assuming rotational symmetry, we
derive center-to-limb profiles for the radial Stokes parameter $Q_{\rm r} (r)/I(r)$.
Because of the scattering symmetry, $Q_{\rm r} (r)/I(r)$ is essentially zero in the disk center, whereas the 
polarization increases for larger radii, reaching a maximum in the seeing halo at $R_{\rm max} > R_{\rm Titan}$.
Depending on wavelength, we measured maximum fractional polarization values 
in the range of $\sim$2-7~\% with the highest value obtained in the $NIC1$ band at 1 $\mu$m.

The observed maximum limb polarization is averaged down by the limited resolution and the PSF structure of HST. 
This has been taken into account in our analysis of the measured disk-integrated radial polarization $\langle Q_{\rm r} /I\rangle_{\rm m}$.
Using synthetic PSF profiles for HST we have modeled the resolution effect on the polarization, and we derive corrected values
for the intrinsic $\langle Q_{\rm r} /I\rangle$ of Titan (Table \ref{intpol}).
We find $\langle Q_{\rm r} /I\rangle\approx 1.2~\%$ in the UV, increasing to about 5.5~\% at \mbox{1 $\mu$m}, and then decreasing again to about 3.4~\% at 2 $\mu$m.
For comparison, a semi-infinite, conservative ($\omega =1$) Rayleigh-scattering atmosphere produces a disk-integrated limb polarization
$\langle Q_{\rm r} /I\rangle \approx 2.75$ \% \citep{Buenzli2009}.

\begin{figure}[t!]
        \centering
        \includegraphics[width=9cm]{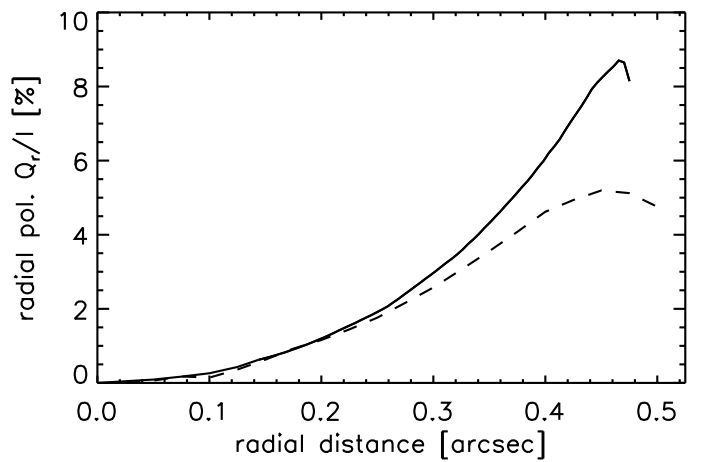}
                 \caption{Measured radial polarization profile for Titan in the $F775W$ band (dashed) versus the modeled profile without PSF smearing (solid).
                 }
                 \label{fig: rprofile model}         
        \end{figure}

Using our radiative transfer model for Titan we predict the maximum limb polarization without PSF smearing, e.g., for observations with larger earth-bound telescopes or
from spacecrafts close to Titan. We find $(Q_{\rm r}/I)^{\rm max}\approx 4.2~\%$ for the $F250W$ band, up to $(Q_{\rm r}/I)^{\rm max}\approx 8.7~\%$ for the $F775W$ and
$(Q_{\rm r}/I)^{\rm max}\approx 10.5~\%$ for the $NIC1$ band, which is even larger than $\approx 8~\%$ for a semi-infinite conservative Rayleigh-scattering atmosphere
\citep{Buenzli2009}. 
The modeled $(Q_{\rm r}/I)^{\rm max}$ for all filters are also given in Table~\ref{intpol}, and Fig.~\ref{fig: rprofile model} 
compares the measured $F775W$ radial polarization profile with the model result. In the figure one can see both the polarization dilution, as well as the decreased resolution 
provided by the HST.

\subsection{Comparison with model calculations}

Our model calculates the intensity and the polarization spectra of the planet at different phase angles $\alpha=0^\circ-180^\circ$, and in case of backscattering ($\alpha=0^\circ$), 
the limb polarization as a function of the radial distance from the disk center. 

Overall, we find a good agreement at the level of $(\Delta A_{\rm g}) / A_{\rm g} \approx 0.1-0.2$ or $(\Delta p) / p \approx 0.1$ respectively between model, literature values, and our observations.
In one filter ($NIC1$ at 1 $\mu$m) we find a discrepancy of $\Delta p/p=0.25$ between the measured limb polarization $\langle Q_{\rm r} /I\rangle$
and our model, for which the cause is still uncertain. Despite this outlier, our analysis
shows that limb-polarization measurements potentially offer an additional diagnostic tool for investigating the properties of scattering particles 
in Titan with earth-bound observations. 
Since our model assumes a rotationally homogeneous atmosphere, our results further show that the locally derived haze and atmosphere parameters from the Huygens probe are indeed representative for Titan. 

Additional observations of Titan with increased resolution and a high polarimetric sensitivity better than 0.1~\%, e.g., using the upcoming SPHERE instrument at the VLT \citep{Beuzit2008}, might reveal structure along the limb 
and temporal changes in the polarization. 
Such limb polarization measurements could be useful for investigating local haze properties of Titan, 
and the impact of short-term and seasonal variations. 
In particular, the limb polarization is very sensitive to the maximum haze single scattering polarization $p_m$, 
which is strongly depending on the monomer size of the small haze aggregates
\citep[e.g.,][]{Tomasko2008}. 
Changing $p_m$ in the $F775W$ band from, e.g., $p_m=0.96$ to $p_m=0.94$,
i.e., corresponding to an increase of the monomer radius of about 15~\% (rough estimate based on Fig.~18 by \citealt{Tomasko2008}), 
reduces the limb polarization by 
$\Delta p / p \approx 0.1$, while the albedo remains essentially unchanged.
  
For other atmospheric parameters like $\tau_{\rm haze}$, $\omega_{\rm haze}$, or the CH$_4$ fraction the geometric albedo $A_{\rm g}$ and the limb polarization
$Q_{\rm r}/I$ will change together ($\omega_{\rm haze}$) or in opposite direction ($\tau_{\rm haze}$). Thus no structure in the limb polarization
is expected if there is no strong albedo feature, like a north-south asymmetry.

\subsection{Prospects for extra-solar planets}

Differential polarimetric imaging is a particularly promising technique for the detection and characterization of extra-solar planets.
With sensitive polarimetry the measurable contrast between star and planet can be enhanced by searching for the polarized signal of the scattered 
light from the planet within the halo of the unpolarized light from the star \citep[e.g.,][]{Schmid2006}. 
The upcoming planet finder instruments SPHERE \citep{Beuzit2008} and GPI \citep{Macintosh2012}
will both provide improved performance for substantial progress in this direction.

The measurable polarization contrast can be described by
\begin{equation}
C_p(\alpha,\lambda)= p(\alpha,\lambda) f(\alpha,\lambda)(R_p/d_p)^2\,,
\label{cpol}
\end{equation}
where $\alpha$ is the phase angle, $R_p$ the radius of the planet, 
$d_p$ its separation to the star, $f(\alpha,\lambda)$ is the phase-dependent reflectivity, and $p(\alpha,\lambda)$ is the
integrated fractional polarization. Therefore, the investigation of
$p(\alpha,\lambda)$ and the polarization flux 
\mbox{$p(\alpha,\lambda) \times f(\alpha,\lambda)$} 
of Titan is important for planning future observing projects on
extrasolar planetary systems and interpreting observational 
data.

On the one hand, Titan shows that atmospheres with thick layers of small aggregate haze particles produce a very strong polarization signal of $p(\alpha\approx 90^\circ)\approx 50~\%$ 
over a wide wavelength range from the UV at 300 nm to the near-IR at 2 $\mu$m.
The bottom panel of Fig.~\ref{fig: results quadrature} gives the expected polarized flux $p\times f$ of Titan if we would observe the object at quadrature phase $\alpha=90^\circ$.
In the optical  $p\times f ~(90^\circ)$ is increasing with wavelength with $p\times f~(90^\circ)\approx 0.02$ at $\lambda=450$ nm, 
up to $p\times f~(90^\circ)\gtrsim 0.03$ at $\lambda=850$ nm. Above 850 nm the polarized flux is strongly decreased in the methane absorption bands. 
Therefore, planets with hazy atmospheres and aerosol properties similar to Titan could be particularly good candidates for detection with ZIMPOL/SPHERE
\citep{Beuzit2008, Schmid2006zimpol} because of their large polarization over the entire wavelength range of the instrument (520-900 nm).

On the other hand, the polarization signal could be strongly reduced for a planet with a Titan-like atmosphere but consisting of larger aggregates. 
If the monomers were about a factor of two larger then the single scattering polarization in the $F775W$ band is reduced by about $\Delta p_m/p_m \approx 0.2$ 
(rough estimate based on \citealt{Tomasko2008}),
and the quadrature polarization will be reduced by $\Delta p/p\approx 0.4$ to about $p(90^\circ)\approx 25~\%$.
The polarization signal of a pure Rayleigh scattering atmosphere would also be different with a high signal in shorter wavelength bands between 550-700 nm, 
and a significantly lower signal for the longer filter bands between 700-800 nm \citep{Buenzli2009}. 
Therefore, a good knowledge of the polarization properties of the hazy atmosphere of Titan is useful for the search and investigation of the polarimetric signal of extra-solar planets.


\begin{acknowledgements} 
We thank the referee for very thoughtful comments and suggestions which lead to a much improved revised version of the paper.
Part of this work was supported by the FINES research fund by a grant through the Swiss National Science Foundation (SNF).
\end{acknowledgements}

\bibliographystyle{aa}    
\bibliography{titan}    

\begin{appendix}
\section{Titan scattering model parameters}\label{appendix a}

\subsection{Number densities and column density}
For the calculation of the methane number density $n_{\rm CH_4}$ and the scale height $Z_{\rm CH_4}$ the following formulas are used:
\begin{equation}\label{eq: number density CH4}
n_{\rm CH_4} = \frac{\rho~[{\rm g/cm}^3] \cdot 6.02 \cdot 10^{23}}{28 / f_{\rm CH_4} - 12}
\end{equation}

\begin{equation}\label{eq: number density N2}
n_{\rm N_2} = n_{\rm CH_4} \cdot \frac{f_{\rm N_2}}{f_{\rm CH_4}} = n_{\rm CH_4} \cdot \frac{1- f_{\rm CH_4}}{f_{\rm CH_4}}
\end{equation}

\noindent{Derivation:}
\begin{itemize}
\item $n_{\rm tot} = \frac{\rho\,[{\rm g/cm}^3]  \cdot \rm N_{\rm A}}{\mu} = n_{\rm CH_4} + n_{\rm N_2}$
\item $\mu = 28\cdot (1-f_{\rm CH_4}) + 16\cdot f_{\rm CH_4}$
\item $n_{\rm CH_4} = f_{\rm CH_4} n_{\rm tot} = \frac{f_{\rm CH_4}\, \rho\,[{\rm g/cm}^3]  \cdot \rm N_{\rm A}}{28(1-f_{\rm CH_4}) + 16 f_{\rm CH_4}}$
\end{itemize}

\noindent{Column density in km-am\footnote{$1~\textnormal{km-am} = 2.687\cdot 10^{24}\,\textnormal{molecules}\,{\rm cm}^{-2}$}:}
\begin{equation}
Z ~[1 / \rm{km^2}] = 10^{15} \int n~[1/\rm{cm^3}]\,\rm d h = 2.687 \cdot 10^{34} \cdot Z ~[\textnormal{km-am}] 
\end{equation}

\subsection{Rayleigh scattering optical depth}
This is from PDS\footnote{http://pds-atmospheres.nmsu.edu/education\_and\_outreach/}:
\begin{equation}
\tau_{\rm ray, sc} = \tau_1(\rm H_2) \cdot (10.1509 \cdot \rm Z_{\rm CH_4} + 4.6035 \cdot \rm Z_{\rm N_2}) \, ,
\end{equation}
with $\rm Z_i \, [\textnormal{km-am}]$ and the optical depth per km-am of hydrogen given as ($\lambda~[\mathring{A}]$):
\begin{align}
\tau_1({\rm H_2}) = 2.687 \cdot \left( \frac{8.14 \cdot 10^{11}}{\lambda^4} + \frac{1.28 \cdot 10^{18}}{\lambda^6} + \frac{1.61 \cdot 10^{24}}{\lambda^8}\right)
\end{align}

\subsection{Methane absorption}
Methane absorption from \citet{Karkoschka2010}. Below 1 $\mu$m these methane absorption coefficients
are generally close to those by \citet{Karkoschka1998},
\begin{align}
{\rm log} \, \kappa (T) =\, & 0.5z(z-1)\,{\rm log}\, \kappa (100) + (1-z^2)\, {\rm log}\, \kappa (198) \notag \\
 &+ 0.5z(z+1)\,{\rm log}\, \kappa (296) \, , 
\end{align}
with $z=(T-198) / 98$; and $\kappa(100)$, $\kappa(198)$, and $\kappa(296)$ are given in Table 4 of the supplementary material of \citet{Karkoschka2010}. 
Therefore, we get
\begin{equation}
\tau_{\rm CH_4} = \kappa_{\rm CH_4} \cdot \rm Z_{\rm CH_4} \, .
\end{equation}

\subsection{Haze optical depth and single scattering albedo}
The haze optical depth $\tau_{\rm h}$ and single scattering albedo $\omega_{\rm haze}$ are taken from \citet{Tomasko2008}. 
 $\tau_{\rm h}$ is given for three altitude regions, above 80 km, 30-80 km, and below 30 km (see Figs. 47, 50). Above 80 km, the cumulative
 optical depth increases with decreasing altitude with a scale height of 65 km. Between 80 and 30 km, and below 30 km the variation is linear with two different slopes. 
 The wavelength dependence for the three regions is taken from Figure 47 of \citet{Tomasko2008}: 
\begin{align}
\tau_{80} (\lambda) &= 1.012 \cdot 10^7 \cdot \lambda^{-2.339} \\
\tau_{30} (\lambda) &= 2.029 \cdot 10^4 \cdot \lambda^{-1.409} \\
\tau_{0} (\lambda) &= 6.270 \cdot 10^2 \cdot \lambda^{-0.9706} \\
\tau_{>80} (h, \lambda) &= \tau_{80}(\lambda) \cdot e^{-\frac{h-80 \, {\rm km}}{65 \, {\rm km}}} \\
\tau_{30-80}(h, \lambda) &= \tau_{80}(\lambda) + \tau_{30}\left(1-\frac{h-30\,{\rm km}}{50\,{\rm km}}\right) \\
\tau_{<30}(h, \lambda) &= \tau_{80}(\lambda) + \tau_{30} + \tau_{0}\left(1-\frac{h}{30\,{\rm km}}\right) \, .
\end{align}

The single scattering albedo $\omega_{\rm haze}$ is given in \citet{Tomasko2008} for three altitude regions (Figure 48, Table 2), above 144 km, 30-80 km, and below 30 km. 
The wavelength dependence of $\omega_{\rm haze}$ for the region above 144 km and 30-80 km is given in Figure 48 of \citet{Tomasko2008} whereas for the region below 30 km
we adopt a three-dimensional polynomial fit to the values given in Table 2 of \citet{Tomasko2008}.
For the regions below 30 km, 30-80 km, and above 144 km, we assume $\omega_{\rm haze}$ to be constant with altitude, whereas for the region 80-144 km we linearly interpolate
between the values at 80 km and 144 km: 
\begin{align}
\omega_{\rm haze, >144} (h, \lambda) &= \omega_{\rm haze, 144} (\lambda) \\
\omega_{\rm haze, 80-144} (h, \lambda) &= \omega_{\rm haze, 30-80} (\lambda) \notag \\
&\,\,\,\,\,+ \frac{\omega_{\rm haze, 144} - \omega_{\rm haze, 30-80}}{64\,{\rm km}} \cdot (h-80\,{\rm km}) \\
\omega_{\rm haze, 30-80} (h, \lambda) &= \omega_{\rm haze, 30-80} (\lambda) \\
\omega_{\rm haze, <30} (h, \lambda) &= \omega_{\rm haze, <30} (\lambda) \, .
\end{align}

\subsection{Effective single scattering albedo}
\begin{align}
\tau_{\rm tot} & = \tau_{\rm ray, sc} + \tau_{\rm CH_4} + \tau_{\rm haze} \\
\omega_{\rm ray} &= \tau_{\rm ray, sc} / (\tau_{\rm ray, sc} + \tau_{\rm CH_4}) \\
\omega_{\rm ray, eff} &= \tau_{\rm ray, sc} / \tau_{\rm tot} \\
\omega_{\rm haze, eff} &= (\omega_{\rm haze} \cdot \tau_{\rm haze}) / \tau_{\rm tot} 
\end{align}

\end{appendix}

\end{document}